# Benchmarking Universal Machine-Learning Interatomic Potentials for Temperature-Dependent Elasticity of Binary and High-Entropy Refractory Carbides

Miroslav Lebeda[a,b], Jan Drahokoupil[a,b], Šimon Svoboda[a], Petr Vlčák[a]

[a] *Faculty of Mechanical Engineering, Czech Technical University in Prague, Technická 4, 16607 Prague 6, Czech Republic*

[b] *FZU – Institute of Physics of the Czech Academy of Sciences, Na Slovance 2, 18200 Prague 8, Czech Republic*

**Corresponding author:** Miroslav Lebeda, lebedmi2@cvut.cz

## Abstract

Accurate predictions of temperature-dependent stiffness are essential for assessing the high-temperature performance of refractory carbides, yet the cost of *ab initio* molecular dynamics (AIMD) limits systematic exploration across compositions and temperatures. Universal machine-learning interatomic potentials (uMLIPs) offer a computationally efficient alternative, but their reliability for finite-temperature elasticity remains insufficiently established. Here, we benchmark nine uMLIPs against consistent AIMD reference data for five binary and two high-entropy (HE) carbides between 300 and 1200 K. Elastic constants and the corresponding bulk, shear, and Young's moduli are obtained using stress-strain molecular dynamics, explicitly sampling thermal atomic motion and anharmonic effects beyond thermal expansion alone. We assess absolute elastic properties and normalized thermal softening separately. MACE-MH-1 achieves the lowest overall mean absolute percentage error (5.7%). MACE-MH-1 and DPA4-Mini reproduce thermal softening most accurately, with mean deviations of 2.5 and 2.3 percentage points, respectively. All models generally underestimate stiffness, with larger equilibrium volumes relative to the AIMD reference likely contributing to this trend for most models. $C_{12}$ exhibits the largest model-dependent errors. The HE carbides are described with accuracy comparable to that of the binary carbides, indicating no apparent accuracy penalty from chemical complexity. Accuracy instead varies with transition-metal composition, with group-V carbides, particularly TaC, presenting the greatest challenge. These results identify promising pretrained models for finite-temperature elasticity in carbides and show why accurate absolute stiffness and thermal softening must be assessed independently.

## 1. Introduction

Refractory transition-metal carbides combine high melting temperatures, hardness, chemical stability, and resistance to mechanical degradation, making them attractive for structural components and protective coatings operating under extreme conditions[1–3]. Their elastic constants govern the response to small deformations and provide fundamental information about mechanical stability, elastic anisotropy, and directional stiffness. Extending binary carbides to multicomponent high-entropy (HE) compositions offers additional possibilities to tailor these properties through chemical disorder[4].

Accurate prediction of the temperature-dependent elastic response is important for understanding these materials and for assessing computational methods intended for their design. Whereas elastic constants can be evaluated relatively efficiently for relaxed structures at 0 K, their temperature dependence presents a more demanding problem. At finite temperature, the elastic stiffness is determined by the strain dependence of the free energy and reflects changes in equilibrium volume as well as vibrational contributions, including intrinsic anharmonic effects beyond the quasiharmonic approximation[5,6]. Static calculations performed at thermally expanded volumes account for only part of this response and may therefore give an incomplete description of thermal softening.

*Ab initio* molecular dynamics (AIMD) provides a direct way for calculating finite-temperature elastic constants by relating imposed strains to time-averaged stresses while explicitly sampling thermally activated atomic motion[7–9]. This approach has successfully described the temperature-dependent elasticity of refractory carbides. For instance, Sangiovanni et al.[10] showed that including atomic vibrations produces thermal-softening trends that are generally more consistent with experimental observations than estimates based on thermal expansion alone. However, the electronic-structure calculations required to sample several strained configurations at multiple temperatures with sufficiently long trajectories make systematic AIMD studies computationally expensive.

Universal (foundation) machine-learning interatomic potentials (uMLIPs) offer a possible way to overcome this limitation. These models are pretrained on large and chemically diverse collections of *ab initio* calculations and can be applied "zero-shot", i.e. without additional fine-tuning, to many compositions without constructing a separate training dataset for every target material[11,12]. A further practical advantage is that many current uMLIPs are openly distributed, with their trained model parameters and implementations available through public repositories. Nevertheless, broad elemental coverage does not guarantee uniform accuracy for every property. Finite-temperature elasticity is particularly demanding because it depends on stresses, potential-energy-surface curvature, and thermally distorted configurations[13].

Recent property-specific benchmarks have evaluated uMLIPs for, e.g., crystal stability, lattice dynamics, surface and defect energetics, and elastic and thermomechanical properties[14–20]. Collectively, these studies demonstrate that small errors in energies and forces near equilibrium do not necessarily guarantee accurate potential-energy-surface curvature, reliable response properties, or robust behavior in configurations and thermodynamic conditions underrepresented in the training. Moreover, finite-temperature MD benchmarks have reported substantial model dependence in predicted bulk moduli and thermal expansion across different material classes[21]. However, the ability of pretrained uMLIPs to reproduce the temperature dependence of the single-crystal elastic tensor remains comparatively underexplored, particularly for refractory carbides and chemically disordered HE ceramics. This motivates the central question addressed in the present work: can broadly pretrained MLIPs reproduce the finite-temperature elastic response of refractory carbides without additional system-specific training?

A practical obstacle to answering this question is the limited availability of directly comparable finite-temperature AIMD elastic data covering multiple compositions. The dataset reported by Sangiovanni et al.[10] provides a particularly suitable reference, comprising rocksalt TiC, ZrC, HfC, VC, and TaC together with the HE carbides (HECs), (Ti,Zr,Hf,Ta,W)C and (V,Nb,Ta,Mo,W)C, at 300, 600, 900, and 1200 K. Because all systems were evaluated using the same computational methodology, the dataset reduces inconsistencies associated with combining reference values from different studies. Its inclusion of both binary and multicomponent carbides also enables examination of whether model accuracy deteriorates as the diversity of local chemical environments increases from binary to multiple components.

In this work, we benchmarked nine uMLIPs, each representing a distinct model architecture, against the finite-temperature AIMD results of Sangiovanni et al[10]. Predictive performance is evaluated for the elastic constants and the derived moduli separately in terms of (i) the absolute stiffness at each investigated temperature and (ii) the normalized temperature dependence relative to 300 K. This distinction identifies models that reproduce the thermal-softening rate despite a systematic offset in stiffness, as well as models that describe the absolute stiffness well but underestimate its decrease with temperature. For the HECs, we additionally assess the sensitivity of the results to the SQS configuration, and we compare the predicted lattice parameters and thermal expansion with the *ab initio* reference. By comparing performance across temperature, elastic quantity and chemical complexity, this study aims to determine the suitability and limitations of current uMLIPs for finite-temperature simulations of refractory carbides.

## 2. Methods

### 2.1 Structural models

The initial rocksalt structures of the binary carbides TiC, ZrC, HfC, VC, and TaC were obtained from the Materials Cloud Three-Dimensional Structure Database (MC3D)[22] and Crystallography Open Database (COD)[23]. Periodic 4 x 4 x 4 supercells containing 512 atoms were constructed for all investigated materials. The HECs, $(Ti_{0.2}Zr_{0.2}Hf_{0.2}Ta_{0.2}W_{0.2})C$ and $(V_{0.2}Nb_{0.2}Ta_{0.2}Mo_{0.2}W_{0.2})C$, were each represented by six independently generated special quasirandom structures (SQSs)[24,25]. Chemical disorder was introduced on the metal sublattice, whereas the carbon sublattice remained fully occupied. Because the 512-atom supercells contain 256 metal sites, each SQS contained 51 atoms of each of the four other metals and 52 W atoms, together with 256 C atoms. The SQSs were generated using the ICET package[26]. Standalone Python scripts for the SQS searches were created using the ICET-SQS interactive interface available at github.com/bracerino/icet-sqs-gui. Each SQS search comprised 5 000 000 simulated-annealing Monte Carlo steps. The cluster space included pair interactions with cutoff distances of up to 5 Å and triplet interactions with cutoff distances of up to 4 Å.

## 2.2 Universal machine-learning interatomic potentials

Nine publicly released pretrained uMLIP models representing different architectures were evaluated (models summarized in **Tab. 1**). The model selection was guided by public availability, reported general predictive performance, and computational feasibility for the MD simulations required by finite-temperature stress-strain workflow. Specifically, each model had to permit simulations of the investigated 512-atom structures on a single GPU with 24 GB memory (RTX 4090) within reasonable computational time. Preference was therefore given to newer models offering a practical balance between accuracy, memory requirements, and simulation throughput. Calculations were partially performed on RTX 4090 24 GB and partially on H100 80 GB GPUs. Among the evaluated models, cuEquivariance acceleration was available for MACE, NequIP, and SevenNet and was enabled for all three. FP32 was used for MACE, SevenNet, and Orb-v3 where supported. Other models retained their default precision. All uMLIPs were applied using their original pretrained parameters without system-specific fine-tuning. The Python scripts for finite-temperature stress-strain workflow were generated using the uMLIP-Interactive interface[27].

| | **SevenNet**[28] | **MACE**[29,30] | **PET/UPET**[31,32] | **GRACE**[33] | **MatterSim**[34] | **Orb**[35] | **NequIP**[36,37] | **DPA**[38] | **ALIGNN-FF**[39,40] |
|---|---|---|---|---|---|---|---|---|---|
| **Model** | 7net-nano-5.5[41] | MACE-MH-1, omat_pbe[42] | PET-MAD-1.5-S[43] | GRACE-2L-OMAT-medium-ft-AM[44] | MatterSim-v1.0.0-1M[34] | orb-v3-conservative-inf-omat[35] | NequIP-OAM-M v0.1[45] | DPA4-Mini[46] | ALIGNN-FF–MatPES-PBE[47] |
| **Training data, *DFT level*** | SevenNet-Omni[48] distillation *PBE(+U)* | OMat24[49] + multi-domain training *PBE(+U)* | OMat24 → MAD-1.5[50] *$r^2$SCAN* | OMat24 → sAlex[51] + MPtrj[52] *PBE(+U)* | MatterSim dataset *PBE(+U)* | OMat24 *PBE(+U)* | OMat24 → sAlex + MPtrj *PBE(+U)* | OMat24 *PBE(+U)* | MatPES-PBE[53] *PBE* |

**Tab. 1:** Architectures and pretrained models of the uMLIPs evaluated in this work. Arrows indicate pretraining followed by fine-tuning. PBE(+U) denotes PBE with Hubbard corrections for selected systems.

## 2.3 Thermal equilibration

All MD simulations were performed through the Atomic Simulation Environment (ASE)[54]. Finite-temperature simulations were performed at 300, 600, 900, and 1200 K using a timestep of 1 fs. At each temperature, the equilibrium simulation cell was established through a 15 ps NPT simulation at zero external pressure. An isotropic Berendsen thermostat and barostat were employed, with temperature and pressure time constants of 0.1 and 1.0 ps, respectively. The averaged cell obtained from the last 10 ps of NPT run was used as the starting cell for the subsequent stress-strain calculations.

### 2.4 Finite-temperature elastic properties calculation

Because all investigated structures possess cubic symmetry (with effective cubic approximation for the disordered HE structures), two independent deformation modes were applied: uniaxial strain along the (*x*) direction and engineering shear strain in the (*yz*) plane. Four strain magnitudes, -1.0%, -0.5%, +0.5%, and +1.0%, were applied to the thermally equilibrated cell. Each strained configuration was equilibrated for 1 ps in the NVT ensemble (not accounted in the stress calculation) and subsequently simulated for a production period of up to 50 ps using a Langevin thermostat with a friction coefficient of 0.02 $fs^{-1}$. Convergence was monitored every 1 ps of production time by re-fitting $C_{11}$, $C_{12}$, $C_{44}$, $B_H$ and $G_H$ and comparing them with the previous check. A run was stopped early once the change in every quantity remained below both 2% and an absolute tolerance of 2 – 3 GPa for 10 consecutive checks, after a minimum production time of 35 ps (otherwise, it continued to the maximum of 50 ps).

The thermodynamic stress tensor, including both virial and kinetic contributions, was sampled every 10 – 20 fs and averaged over the production trajectory. To verify that this difference does not affect the results, the 10 fs series of the other models were resampled at 20 fs, which changed the elastic constants by less than 0.3 GPa. To account for temporal correlations in the fluctuating MD stresses, each production trajectory was divided into five equal time blocks. The stress was averaged separately within each block, and the variation among the block averages was used to estimate the standard error of the overall mean stress. The elastic constants were obtained through weighted linear regression of the time-averaged stresses against the applied strain. The longitudinal stress response provided $C_{11}$, the two symmetry-equivalent transverse responses were averaged to obtain $C_{12}$, and the shear response provided $C_{44}$. The resulting elastic constants were used to calculate the bulk, shear, and Young's moduli using the Voigt-Reuss-Hill (VRH) scheme.

### 2.5 Computational performance

The computational performance of the uMLIPs was evaluated using 20 ps NPT simulations of a TiC supercell at 300 K and zero external pressure on NVIDIA RTX 4090 (24 GB) and H100 (80 GB) GPUs. All benchmarks used the same ASE-based script and the simulation settings described in **Section 2.3**. After discarding the first 1000 steps as warm-up, the average simulation speed was calculated as the total number of MD steps divided by their elapsed wall-clock time. Variability was expressed as the standard deviation of the speeds every 100 steps. GPU memory is reported as the median of the values recorded every 100 steps.

## 3. Results and discussion

We benchmarked five binary carbides (TiC, ZrC, HfC, VC, and TaC) and two high-entropy carbides (HECs), (Ti,Zr,Hf,Ta,W)C and (V,Nb,Ta,Mo,W)C, using nine uMLIPs: MACE-MH-1, PET-MAD-1.5-S, MatterSim-v1.0.0-1M, DPA4-Mini, GRACE-2L-OMAT-medium-ft-AM, NequIP-OAM-M, Orb-v3, SevenNet-Nano 5.5 Å, and ALIGNN-FF MatPES-PBE. At 300, 600, 900, and 1200 K, MD was used to obtain thermally equilibrated lattice parameters and to determine elastic constants and derived moduli from the stress-strain response. The simulations sample thermal atomic motion and its effects on expansion and elasticity. Predictions were compared with published AIMD reference data from Sangiovanni et al[10]. The benchmark workflow overview is summarized schematically in **Fig. 1**.

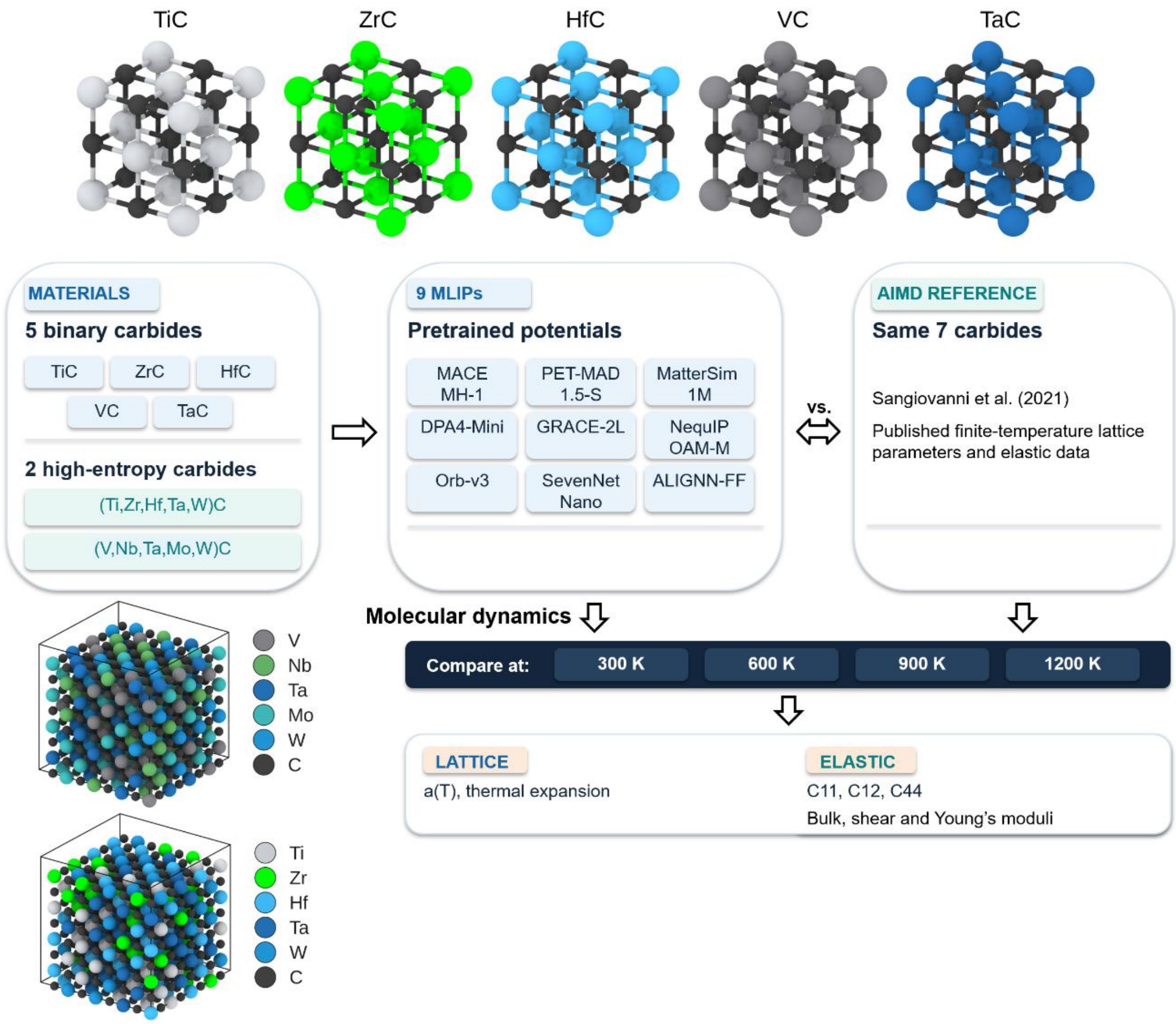

**Fig. 1:** Overview of the finite-temperature benchmark. Nine uMLIPs were applied to five binary and two high-entropy carbides. Molecular dynamics predictions of lattice parameters, thermal expansion, elastic constants, and derived moduli at 300 – 1200 K were compared with published AIMD reference data from Sangiovanni et al[10].

## 3.1 Sensitivity to the SQS configuration

First, we assessed how sensitive the predicted elastic properties of the HE carbides are to the particular arrangement of metal atoms in the 512-atom supercell. Each uMLIP was evaluated on six independent SQS configurations for both, (Ti,Zr,Hf,Ta,W)C and (V,Nb,Ta,Mo,W)C. For each model and property, the standard deviation across the six SQS configurations was calculated at each temperature and then averaged over 300 – 1200 K (**Fig. 2**). For both carbides, the variation is small: the median standard deviation over all models and properties is 0.53 % for (Ti,Zr,Hf,Ta,W)C and 0.59 % for (V,Nb,Ta,Mo,W)C, and more than 90 % of all values lie below 1 %. $C_{11}$ shows the largest variation (on average 0.9 %, about 4.5 – 4.6 GPa), whereas the shear constant $C_{44}$ is the least sensitive (0.3 – 0.4 %, about 0.4 – 0.5 GPa). The Hill moduli vary by about 0.5 – 0.6 % (1.5 – 1.6 GPa for $B_H$, 0.8 GPa for $G_H$ and 2.0 GPa for $E_H$). The largest single value, 1.3 % (6.6 GPa) for $C_{11}$ of (V,Nb,Ta,Mo,W)C predicted by MatterSim-v1.0.0-1M, remains small. All uMLIPs show a comparable degree of SQS-to-SQS scatter, so none is noticeably more sensitive to the local chemical environment than the others. For the subsequent comparison with the DFT reference data, the averaged values from the 6 SQSs were therefore used.

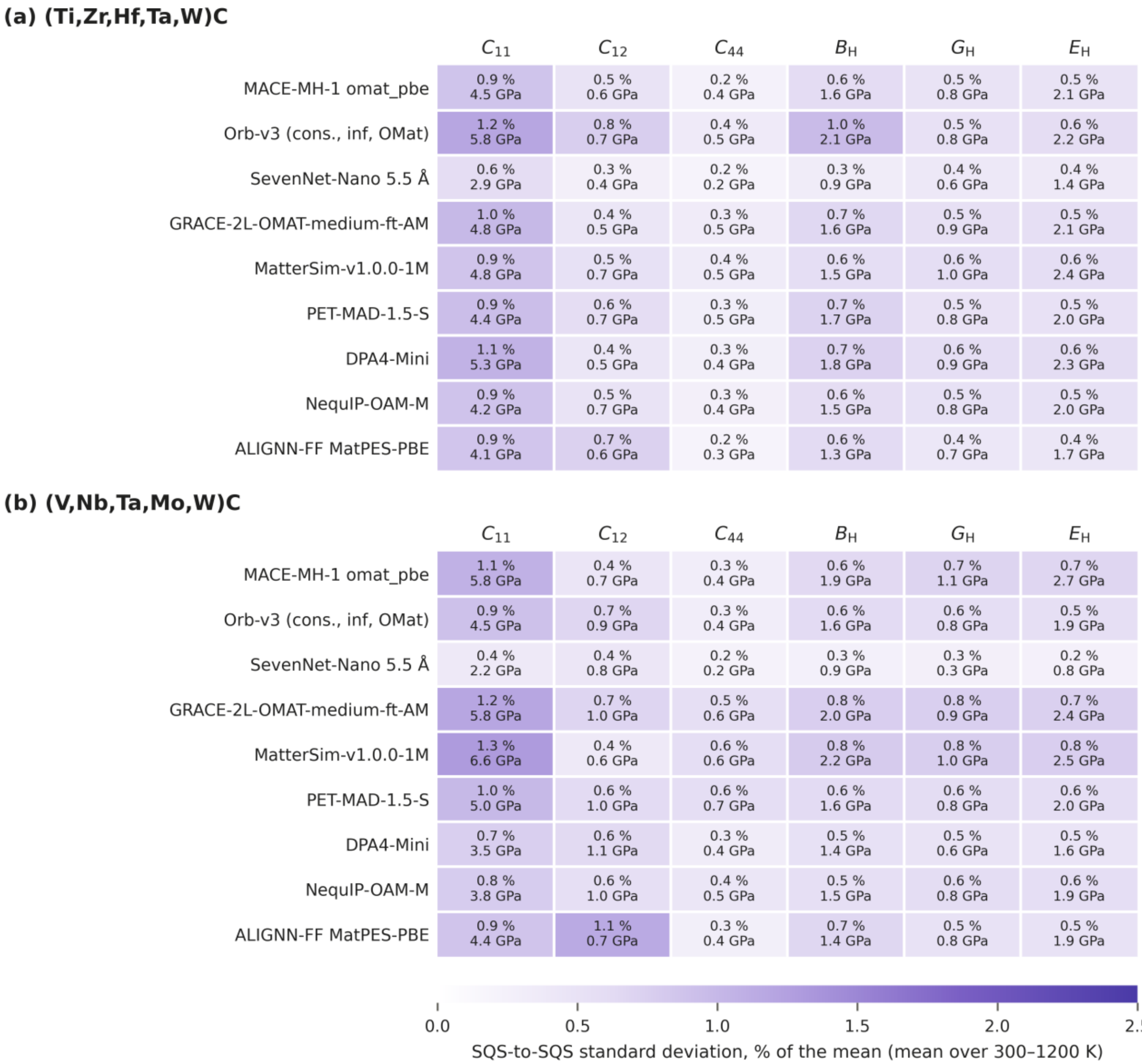

**Fig. 2:** Sensitivity of the uMLIP finite-temperature elastic properties to the SQS configuration for **(a)** (Ti,Zr,Hf,Ta,W)C and **(b)** (V,Nb,Ta,Mo,W)C. Each cell gives the standard deviation across six 512-atom SQS configurations, as a percentage of the mean (top value, with colour scale) and in GPa (bottom). The standard deviation was calculated at each temperature and then averaged over 300, 600, 900 and 1200 K.

### 3.2 Lattice parameters and thermal expansion

As an initial assessment of the uMLIPs at finite temperatures, we compared the lattice parameters obtained from NPT equilibration with the AIMD reference data of Sangiovanni et al.[10] over the temperature range of 300 – 1200 K (**Fig. 3**). The lattice parameter was calculated from the time-averaged NPT cell. For the HE carbides, the resulting lattice parameters were further averaged over the SQS configurations. Static (0 K) PBE lattice parameters from the Open Quantum Materials Database (OQMD)[55] and Yu et al.[56,57] are also included in the comparison.

All uMLIPs reproduce the increase in lattice parameter with temperature. Between 300 and 1200 K, the AIMD lattice parameters increase by 0.6 – 0.8%, corresponding to average linear thermal expansion coefficients of 7.0 – 8.9 × $10^{-6}$ $K^{-1}$. The uMLIPs predict similar thermal expansion, typically approximately 10% greater than AIMD. PET-MAD-1.5-S and MatterSim-v1.0.0-1M show the closest agreement with the AIMD thermal expansion, whereas Orb-v3 and ALIGNN-FF overestimate it by approximately 20% and 40%, respectively, and SevenNet-Nano underestimates it by approximately 15% (the only model that underestimates it).

The absolute lattice parameters exceed the AIMD values for all models, materials, and temperatures, except ALIGNN-FF for TaC (300 – 600 K) and (V,Nb,Ta,Mo,W)C (300 – 900 K). At 300 K, the deviations range from -0.7% to +1.6%, averaging approximately +0.9%, with similar offsets for the binary and HE carbides. Comparison with the static PBE references suggests that this systematic difference largely reflects an offset in the AIMD reference volumes. The three independent static PBE results agree closely, giving lattice parameters of 4.33 - 4.34 Å for TiC, 4.72 - 4.73 Å for ZrC, 4.64 - 4.65 Å for HfC, 4.15 - 4.16 Å for VC, and 4.47 - 4.48 Å for TaC. These values exceed the AIMD results at 300 K by 0.4 - 1.2%, despite the expected increase in lattice parameter due to thermal expansion. By comparison, the uMLIP lattice parameters at 300 K exceed the static PBE references by only 0.2% on average, consistent with a small expansion between 0 and 300 K. MACE-MH-1, ALIGNN-FF, and PET-MAD-1.5-S are closest to the static references, with average deviations of -0.05%, -0.05%, and +0.07%, respectively.

These comparisons indicate that the uMLIP equilibrium volumes are broadly consistent with the static PBE results, while the AIMD reference volumes are systematically smaller. The Γ-point-only Brillouin-zone sampling and lower 300 eV plane-wave cutoff used in the reference AIMD simulations are possible contributing factors, although their effects would require explicit convergence tests to establish. The comparable offsets observed for the two HE carbides suggest a similar origin, but no static reference data are available to verify this interpretation. This systematic volume difference also affects the comparison of elastic properties. A 1% increase in lattice parameter corresponds to an approximately 3% increase in volume. Differences in equilibrium volume may therefore account for part of the lower elastic stiffness predicted by the uMLIPs relative to AIMD.

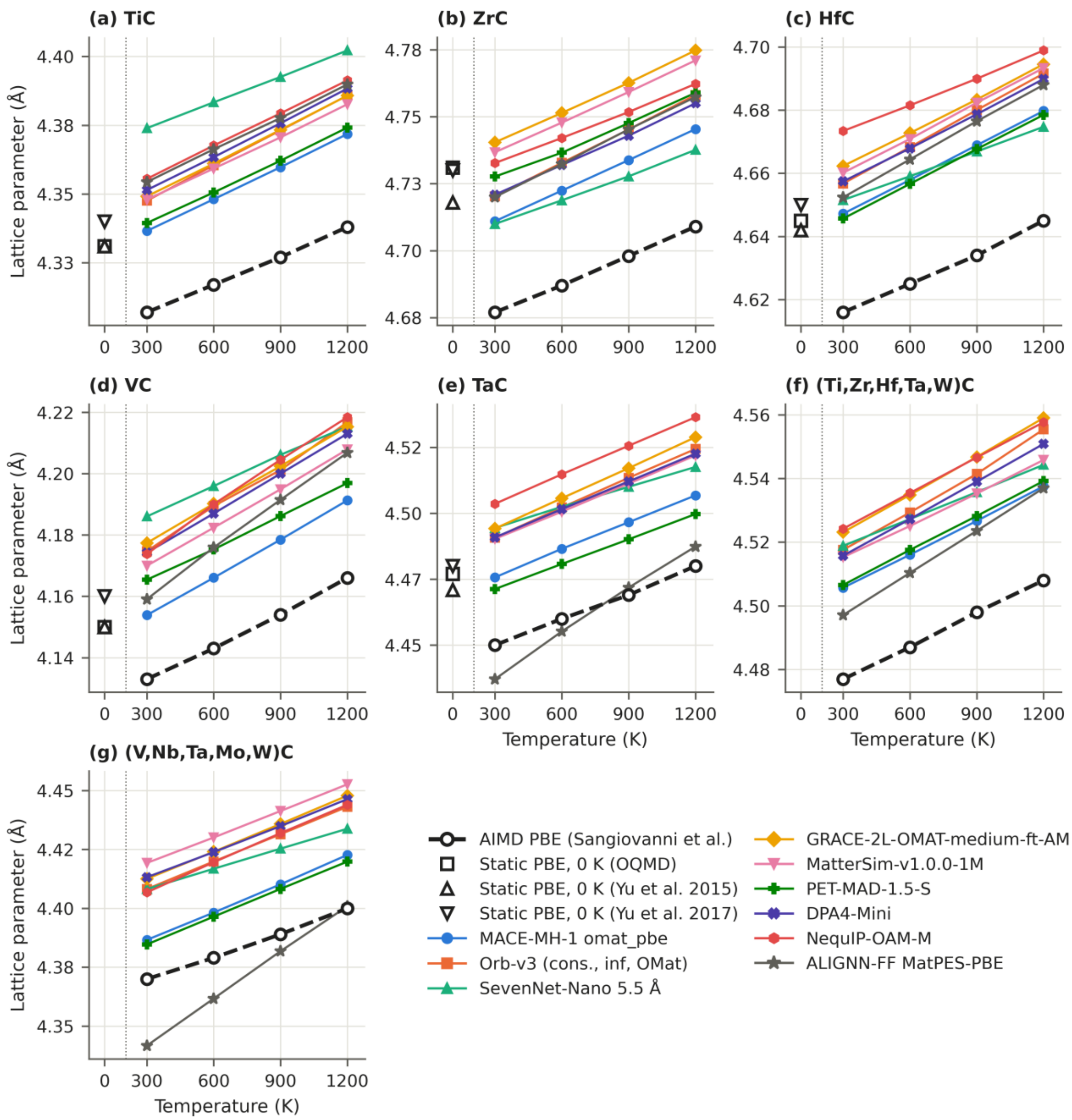

**Fig. 3:** Lattice parameters as a function of temperature obtained from NPT-equilibrated uMLIP cells. Dashed lines show the PBE AIMD reference data of Sangiovanni et al.[10]. For the binary carbides, static PBE lattice parameters at 0 K from OQMD[55] and Yu et al.[56,57] are also shown for the comparison.

### 3.3 Elastic constants and moduli

The elastic constants $C_{11}$, $C_{12}$ and $C_{44}$ and the Hill-averaged bulk, shear and Young's moduli predicted by the uMLIPs at 300 K with the AIMD reference for all 7 carbides are compared in **Fig. 4**. **Supplementary Information** shows the data for the remaining 600, 900, and 1200 K temperatures (**Figs. S1, S2, S3**). Most uMLIPs systematically underestimate the stiffness. $C_{11}$ is lower than the AIMD reference in 97 % of the model-material combinations, with average deviations of -6 to -20 % depending on the carbide. Similarly, the Young's modulus $E_H$ is underestimated in 94% of cases. These deviations are consistent with the larger equilibrium volumes reported in the previous section, suggesting that differences in equilibrium volume contribute to the lower predicted stiffness.

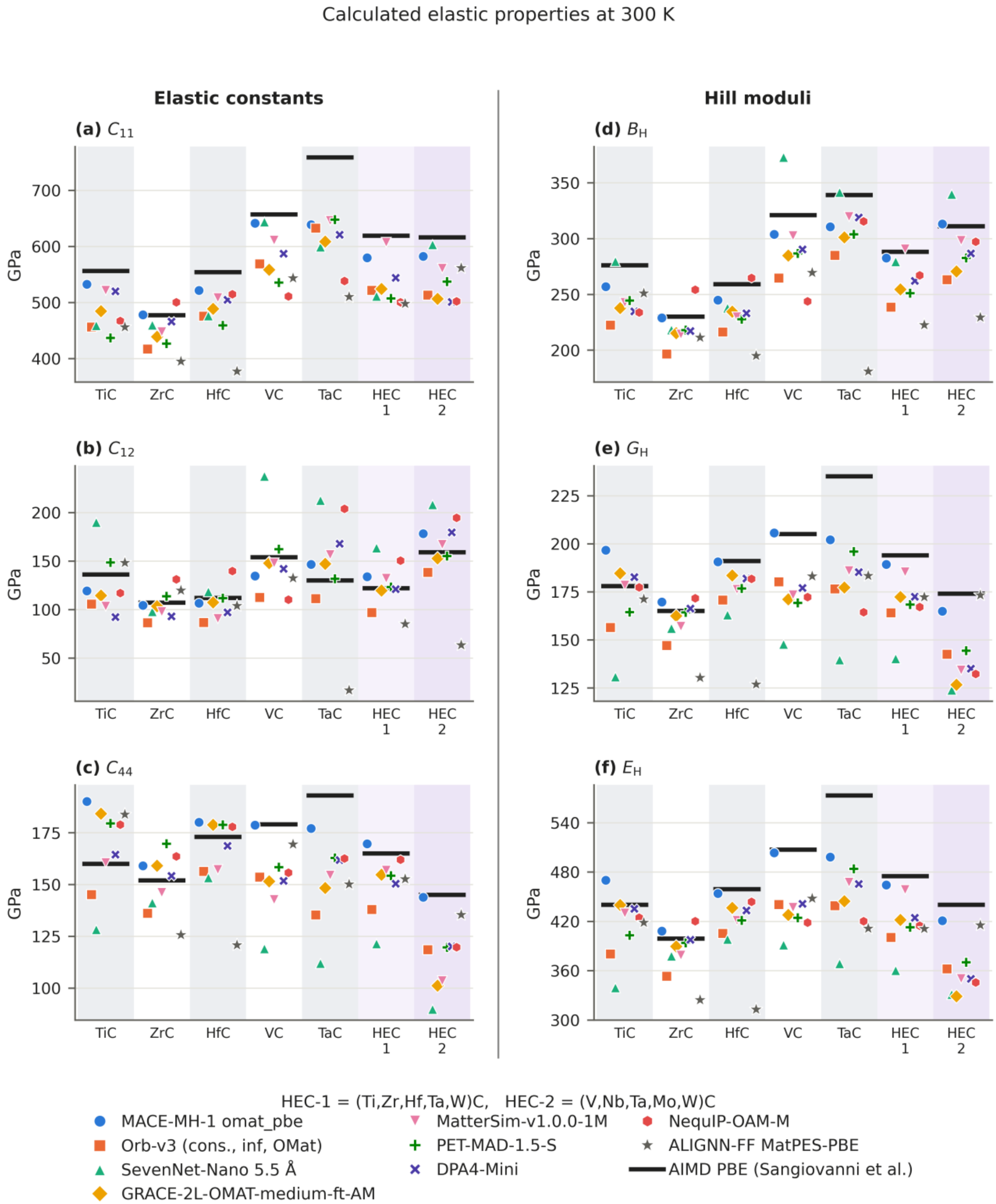

**Fig. 4:** Elastic constants $C_{11}$, $C_{12}$ and $C_{44}$ and Hill-averaged bulk ($B_H$), shear ($G_H$) and Young's ($E_H$) moduli of the 7 carbides at 300 K from the uMLIPs (symbols) compared with AIMD (black lines, Sangiovanni et al[10]).

The agreement is less uniform for the shear response. For the group-IV carbides TiC and ZrC, the predicted $C_{44}$ values lie on both sides of the AIMD reference, with average deviations of +5 % and -1 %, respectively. The corresponding $G_H$ and $E_H$ values agree with AIMD within a few percent on average. In contrast, all models underestimate $C_{44}$ for the group-V carbides VC and TaC and for (V,Nb,Ta,Mo,W)C, with average deviations of -14 to -21 %, resulting in shear moduli that are 14 – 24 % below the reference. TaC exhibits particularly large discrepancies: all models underestimate $C_{11}$ by 15 – 33 % and $G_H$ by 14 – 41 %. The HE carbides show trends consistent with their constituent metals. The average deviation of -8 % in $C_{44}$ is smaller for (Ti,Zr,Hf,Ta,W)C, in which group-IV metals constitute the majority of the metal sublattice, than for (V,Nb,Ta,Mo,W)C (-19 % average deviation).

$C_{12}$ has the relative deviations range from -87 % to +63 %, representing the largest spread among the evaluated properties. SevenNet-Nano overestimates $C_{12}$ for six of the seven carbides (all except ZrC), by up to 63 % for TaC and 54 % for VC (237 GPa compared with 154 GPa). Together with its underestimation of $C_{44}$, this results in the lowest predicted shear moduli among the models. NequIP-OAM-M also overestimates $C_{12}$ for most carbides, whereas Orb-v3 consistently underestimates it. ALIGNN-FF strongly underestimates $C_{12}$ of TaC (17 GPa compared with 130 GPa, -87 %).

Among the evaluated models, MACE-MH-1 shows the closest overall agreement with AIMD at 300 K, with 79 % of its predicted elastic constants and moduli falling within 10 % of the reference, followed by MatterSim-v1.0.0-1M at 64 % falling within 10 % range of the reference. Orb-v3 underestimates nearly all evaluated properties across the seven carbides, with only 5 % of its predictions within 10 % of AIMD. The overall model accuracy across materials and the temperature dependence are discussed in the following section.

### 3.4 Model accuracy and temperature dependence

To compare the uMLIPs quantitatively, the deviations from AIMD were condensed into two measures (**Fig. 5**). Absolute accuracy was assessed by the mean absolute percentage error (MAPE) of each elastic constant and Hill modulus, averaged over all carbides and temperatures from 300 to 1200 K. Temperature dependence was assessed by the softening error, defined as the difference, in percentage points (pp), between the uMLIP and AIMD relative changes [$X$(1200 K) − $X$(300 K)]/$X$(300 K) of each property.

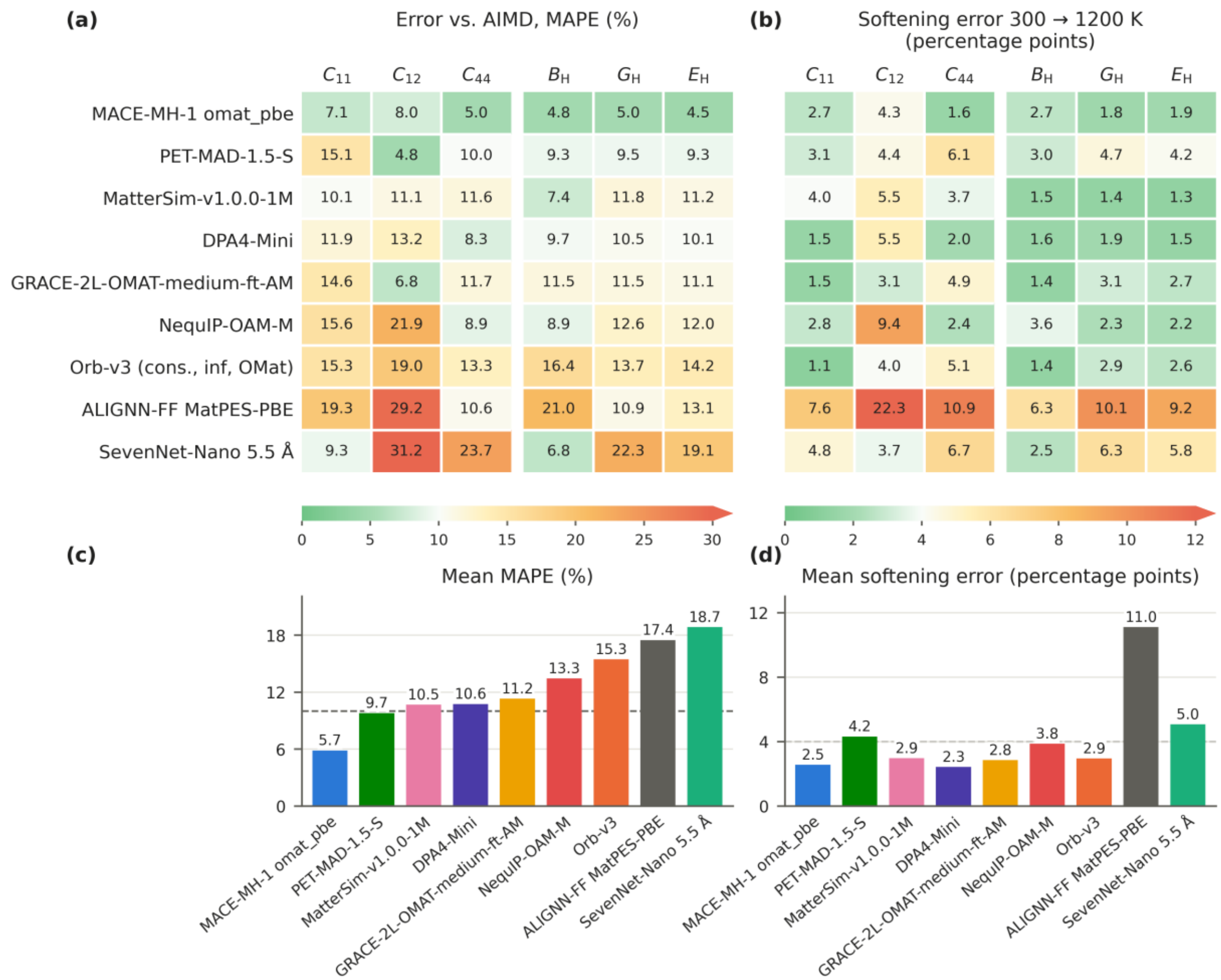

**Fig. 5:** Overall performance of the uMLIPs across TiC, ZrC, HfC, VC, TaC, (Ti,Zr,Hf,Ta,W)C, and (V,Nb,Ta,Mo,W)C. **(a)** Mean absolute percentage errors (MAPEs) of the elastic constants and Hill moduli relative to AIMD (Sangiovanni et al[10]), averaged over all carbides and temperatures from 300 to 1200 K. **(b)** Softening errors, expressed as differences in percentage points between the uMLIP and AIMD relative changes from 300 to 1200 K, averaged over the carbides. **(c)** mean MAPE (%). **(d)** mean absolute softening error in percentage points.

MACE-MH-1 gives the lowest overall MAPE (5.7%) and is the only model with errors below 10% for every property (**Fig. 5, a, c**). It is followed by PET-MAD-1.5-S (9.7%), MatterSim-v1.0.0-1M (10.5%), DPA4-Mini (10.6%), and GRACE-2L-OMAT-medium-ft-AM (11.2%). Larger errors are obtained with NequIP-OAM-M (13.3%), Orb-v3 (15.3%), ALIGNN-FF (17.4%), and SevenNet-Nano (18.7%). Accuracy varies considerably between properties. $C_{12}$ shows the largest spread across models, with MAPE values ranging from 4.8% for PET-MAD-1.5-S to 31.2% for SevenNet-Nano. Almost all models underestimate $C_{11}$ by 7 – 19%. Errors in the Hill moduli are generally smaller because deviations in the individual elastic constants partly cancel (**Fig. 6**). This reduction is most pronounced for SevenNet-Nano (21.4% for elastic constants versus 16.1% for moduli), ALIGNN-FF (19.7% versus 15.0%), and NequIP-OAM-M (15.5% versus 11.2%), but is absent for GRACE-2L-OMAT-medium-ft-AM (11.0% versus 11.3%). Agreement in the moduli therefore does not necessarily imply accurate individual elastic constants.

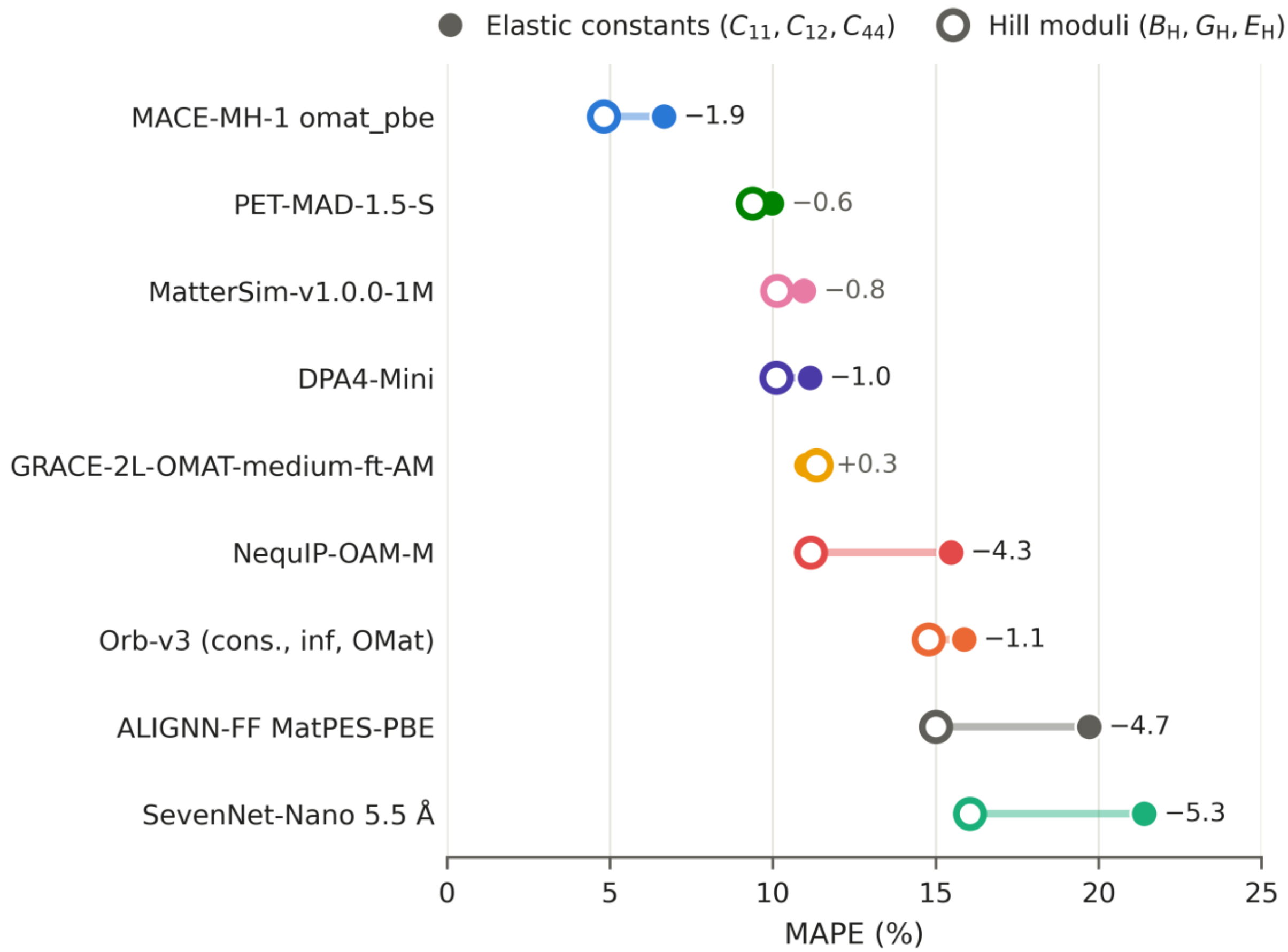

**Fig. 6:** Mean absolute percentage error (MAPE) of the elastic constants ($C_{11}$, $C_{12}$, $C_{44}$, filled circles) and of the Hill moduli ($B_H$, $G_H$, $E_H$, open circles) for each uMLIP, averaged over all carbides and over 300 – 1200 K. The numbers give the change from elastic constants to moduli in percentage points (negative values indicate partial cancellation of errors in the averages).

In AIMD, the shear modulus decreases by 10 – 16 % between 300 and 1200 K, depending on the carbide. MatterSim-v1.0.0-1M, DPA4-Mini and MACE-MH-1 reproduce this softening closely, with softening errors for $G_H$ of only 1.4 – 1.9 pp (**Fig. 5, b**), whereas PET-MAD-1.5-S (4.7 pp), SevenNet-Nano (6.3 pp), and ALIGNN-FF (10.1 pp) predict a too weak reduction of the shear modulus, which decreases on average by only about 8 %, 6 %, and 4 %, respectively. $C_{12}$ softening is the most difficult to reproduce, with errors of 22.3 pp for ALIGNN-FF and 9.4 pp for NequIP-OAM-M. The ALIGNN-FF value is dominated by TaC, for which $C_{12}$ increases from 17 to 36 GPa between 300 and 1200 K. Absolute accuracy and thermal softening are therefore distinct: Orb-v3 combines a large MAPE (15.3%) with a small softening error (2.9 pp), whereas PET-MAD-1.5-S gives relatively accurate property values (MAPE 9.7 %) but underestimates their temperature dependence (softening error 4.2 pp).

Accuracy generally improves slightly with temperature (**Fig. 7**). Averaged over models evaluated for all carbides, the MAPE decreases from 13.4% at 300 K to 11.4% at 1200 K. MACE-MH-1 remains within 5.5 – 6.0%, while larger improvements occur for ALIGNN-FF, NequIP-OAM-M, PET-MAD-1.5-S, and SevenNet-Nano. This improvement partly reflects compensation between underestimated stiffness and insufficient thermal softening: models that soften less than AIMD approach the reference values as temperature increases. Accordingly, the average relative deviation from AIMD changes from -9.5 % at 300 K to -7.8 % at 1200 K.

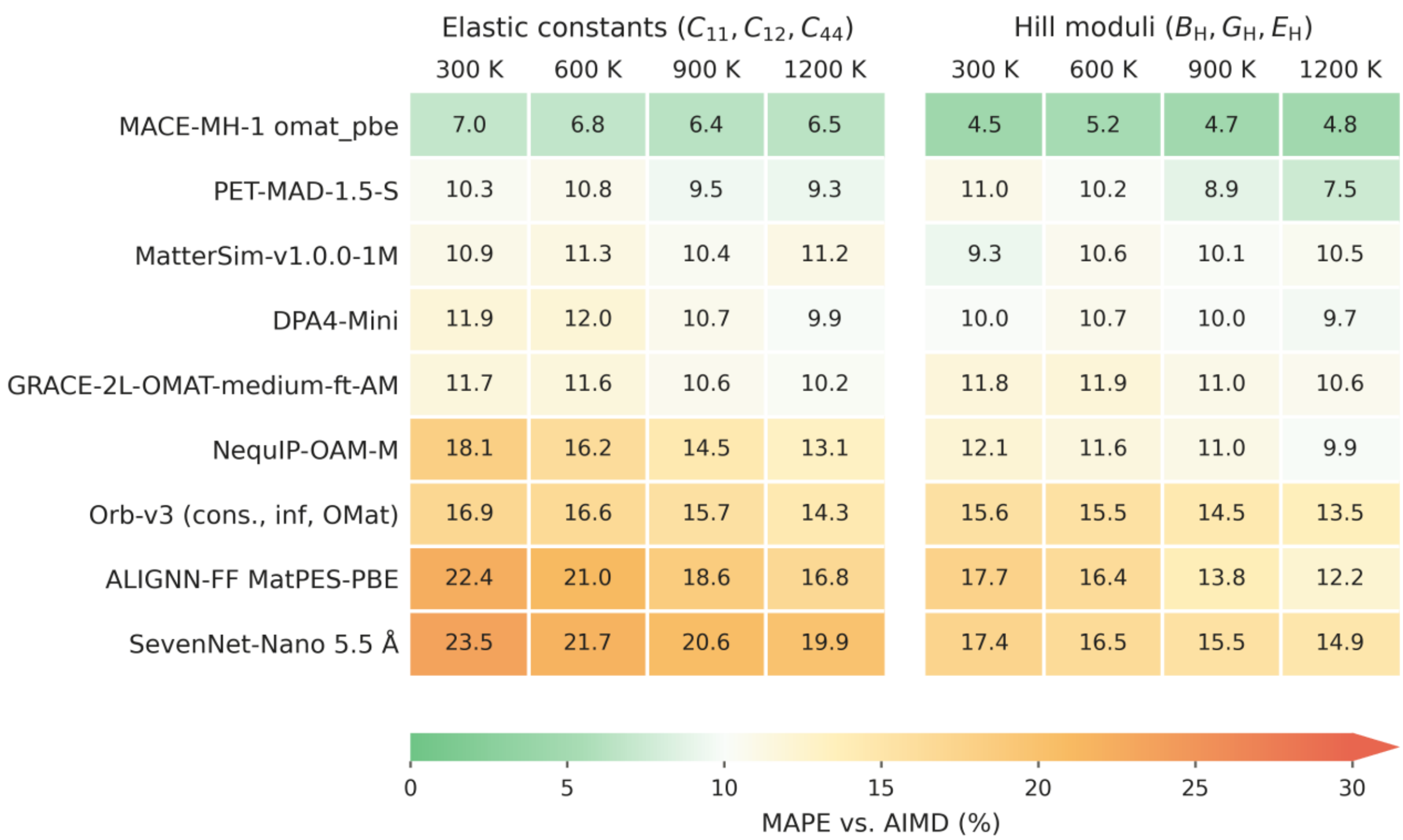

**Fig. 7:** Mean absolute percentage error (MAPE) of the elastic constants ($C_{11}$, $C_{12}$, $C_{44}$) and the Hill moduli ($B_H$, $G_H$, $E_H$) at each temperature, averaged over the carbides.

Overall, MACE-MH-1 provides the best combined description of absolute values and thermal softening. MatterSim-v1.0.0-1M, DPA4-Mini, and GRACE-2L-OMAT-medium-ft-AM also capture the temperature dependence well, with larger absolute errors of approximately 10 – 11%. The temperature-dependent elastic constants and derived moduli for all seven carbides are provided in the Supplementary Information (**Figs. S4, S5**).

### 3.5 Accuracy across materials: binary vs high-entropy carbides

**Fig. 8** compares the absolute accuracy (**Fig. 8, a**) and thermal softening (**Fig. 8, b**) predicted by the uMLIPs for the 7 carbides. Bars show the averages over the 9 models evaluated for all carbides, while symbols represent individual models. This comparison assesses whether the chemically complex HE carbides are more difficult to describe than the binary carbides and if certain carbides are predicted better than others.

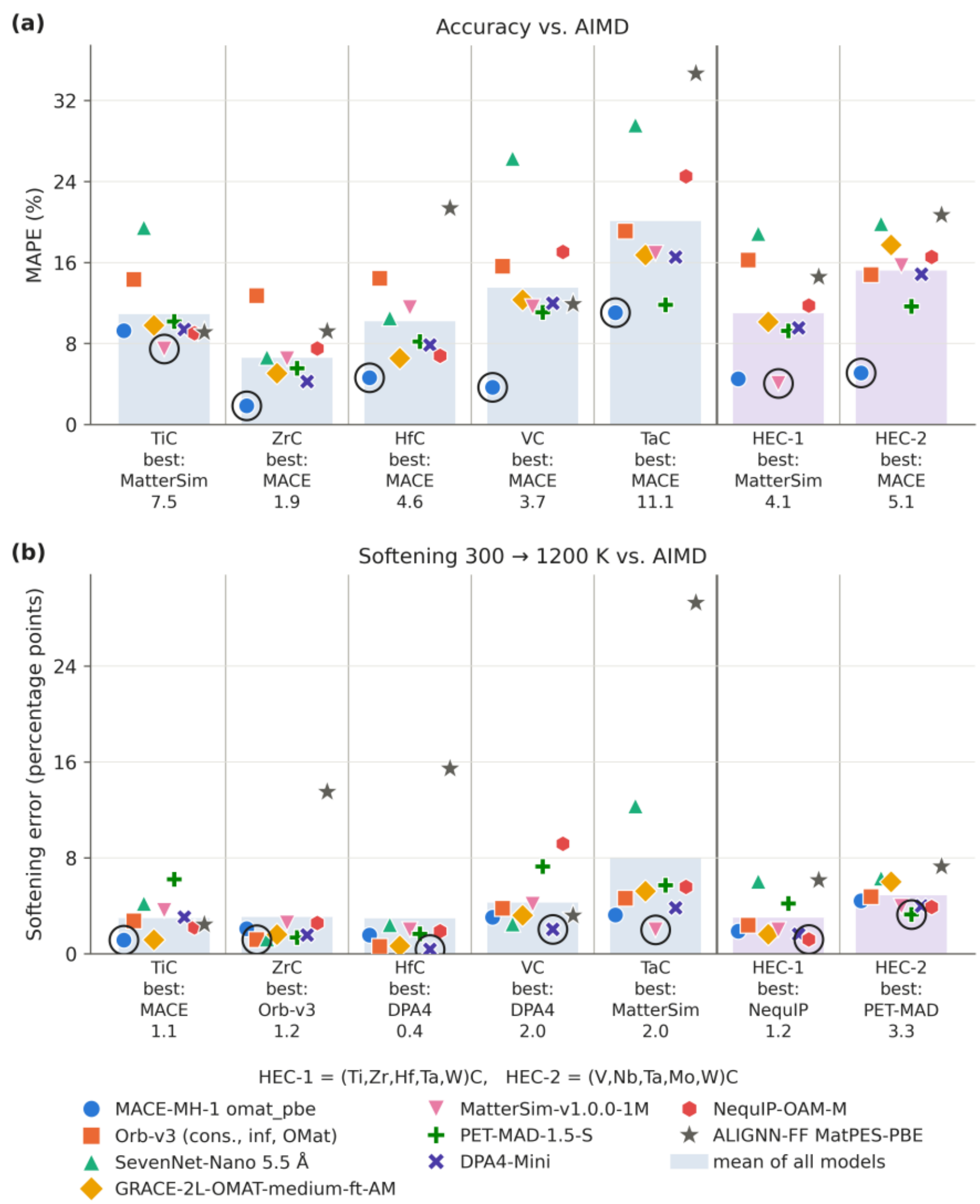

**Fig. 8:** Accuracy of the uMLIPs for each carbide. **(a)** MAPE of C11, C12, C44, B_H, G_H and E_H with respect to AIMD (Sangiovanni et al.[10]) over 300 – 1200 K; **(b)** mean softening error between 300 and 1200 K (percentage points). Symbols: individual models; bars: mean over the nine models evaluated for all carbides; circles: best model per carbide. The numbers below each material give the best-model value.

We found that the HE carbides are described with comparable accuracy to the binary carbides. Their mean errors are 11.0% for (Ti,Zr,Hf,Ta,W)C and 15.2% for (V,Nb,Ta,Mo,W)C, being within the binary range of 6.6 – 20.1%. Their mean softening errors, 3.0 and 4.9 pp, respectively, also fall within the binary range of 3.0 – 8.0 pp. The best models achieve errors of approximately 4 – 5% for both HE carbides: MatterSim-v1.0.0-1M gives 4.1% for (Ti,Zr,Hf,Ta,W)C, and MACE-MH-1 gives 5.1% for (V,Nb,Ta,Mo,W)C. Chemical disorder therefore does not appear to introduce an additional accuracy reduction for these systems (within the measure of comparing to the ab initio reference), consistent with the weak sensitivity to SQS configuration reported in **Section 3.1**.

Instead, accuracy varies systematically with the constituent transition metals. The lowest mean errors are obtained for the group-IV carbides ZrC (6.6%), HfC (10.2%), and TiC (10.9%). Errors are larger for the group-V carbides VC (13.5%) and TaC (20.1%). Only MACE-MH-1 achieves an error below 10% for VC (3.7%), while no model meets this threshold for TaC. Thermal softening follows the same trend, with mean errors of 3.0 – 3.1 pp for the group-IV carbides and 4.3 – 8.0 pp for VC and TaC. The HE carbides broadly follow their constituent metals: (Ti,Zr,Hf,Ta,W)C, containing predominantly group-IV metals, has a similar error to TiC (11.0% versus 10.9%), whereas (V,Nb,Ta,Mo,W)C, containing group-V and group-VI metals, resembles VC (15.2% versus 13.5%). For the latter, MACE-MH-1 is again the only model below 10%.

The larger errors for the group-V-based carbides arise mainly from the shear response. Averaged over 300 – 1200 K, $C_{44}$ is reproduced within ±6% for TiC, ZrC, and HfC, but is underestimated by 13 - 18% for VC, TaC, and (V,Nb,Ta,Mo,W)C. The corresponding Hill shear modulus, $G_H$, is underestimated by 14 – 21 %. TaC is the most difficult case, with an error of 11.1% even for the most accurate model, MACE-MH-1.

MACE-MH-1 gives the lowest absolute error for five of the seven carbides, while MatterSim-v1.0.0-1M performs best for the remaining two, TiC and (Ti,Zr,Hf,Ta,W)C. The model that best reproduces thermal softening varies between carbides, with MACE-MH-1, Orb-v3, DPA4-Mini, MatterSim-v1.0.0-1M, NequIP-OAM-M, and PET-MAD-1.5-S each performing best for at least one material. This variation reinforces the distinction between absolute accuracy and temperature dependence discussed in **Section 3.4**.

### 3.6 Computational performance of the uMLIPs

**Fig. 9** compares the computational performance of the uMLIPs for NPT MD simulations of a 512-atom TiC supercell, using single precision (FP32) and cuEquivariance acceleration where supported. SevenNet-Nano 5.5 Å is the fastest model, achieving 23.4 and 30.5 steps $s^{-1}$ on the RTX 4090 and H100, respectively, approximately twice the throughput of the next-fastest model. PET-MAD-1.5-S, DPA4-Mini, NequIP-OAM-M, MACE-MH-1, Orb-v3 and GRACE-2L-OMAT-medium-ft-AM exhibit broadly comparable speeds, ranging from 8 to 11.5 steps $s^{-1}$ on the RTX 4090 and from 10 to 16 steps $s^{-1}$ on the H100. MatterSim-v1.0.0-1M and ALIGNN-FF MatPES-PBE are the slowest models, reaching 5.9 and 5.2 steps $s^{-1}$ on the RTX 4090 and 7.7 and 7.2 steps $s^{-1}$ on the H100, respectively. The H100 provides a speedup by a factor of ~1.1–1.7 (~1.1 for MACE-MH-1 to ~1.7 for Orb-v3). The RTX 4090 thus achieves approximately 60 – 90 % of the H100 throughput, making it a practical and significantly cheaper alternative for simulations of this system size under the tested settings. GPU memory requirements are also moderate, ranging from 1.1 GiB for SevenNet-Nano to 5.5 GiB for ALIGNN-FF.

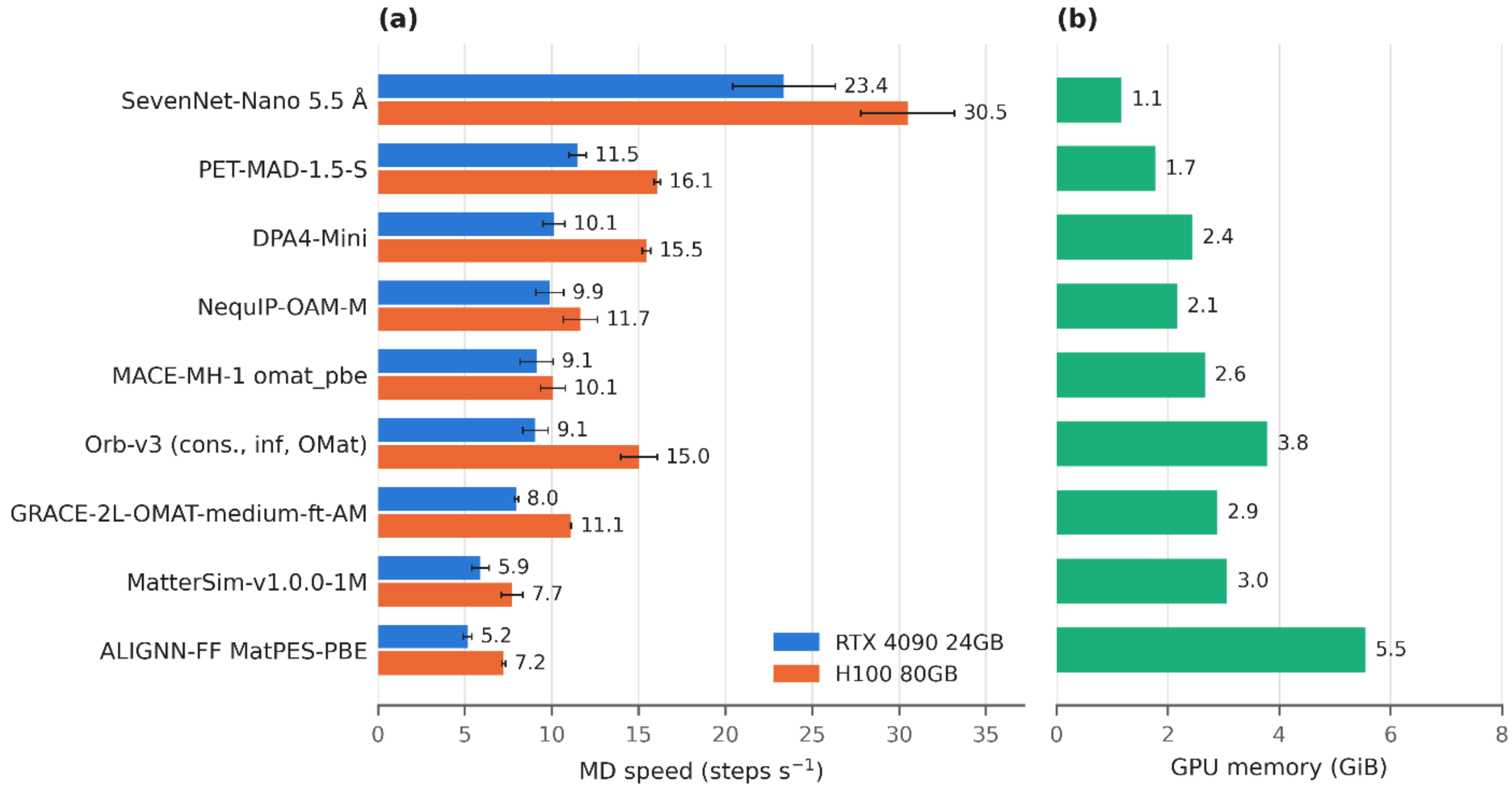

**Fig. 9.** Performance of the uMLIPs for NPT MD of a 512-atom TiC supercell (300 K, 1 fs time step, averaged from 20 000 steps, FP32 and cuEquivariance acceleration where available). **(a)** Simulation speed on an NVIDIA RTX 4090 24GB and an NVIDIA H100 80GB, given as averaged MD steps per second with standard deviation. **(b)** Used GPU memory on H100 as a whole-device memory medians during the calculation for each model.

## 4. Conclusions

We benchmarked nine pretrained universal machine-learned interatomic potentials for the finite-temperature elastic properties of five binary carbides (TiC, ZrC, HfC, VC, and TaC) and two high-entropy carbides, (Ti,Zr,Hf,Ta,W)C and (V,Nb,Ta,Mo,W)C, between 300 and 1200 K, using explicit stress-strain molecular dynamics and AIMD reference data. All models reproduce the thermal expansion, and most models reproduce the overall softening trend, while mean benchmark accuracy generally improves with temperature. ALIGNN-FF shows a softening error of 11.0 pp, more than twice that of any other model. Most models underestimate the absolute stiffness, partly reflecting equilibrium lattice parameters approximately 1% larger than the AIMD values. Comparison with static PBE calculations suggests that this difference largely reflects smaller AIMD reference volumes. MACE-MH-1 gives the lowest mean error (5.7%) and remains below 10% for every property, followed by PET-MAD-1.5-S (9.7%) and several models with errors of 10 – 11%. SevenNet-Nano is the fastest model but has the largest mean error (18.7%). Partial cancellation of errors in the Hill averages means that accurate polycrystalline moduli do not necessarily imply accurate single-crystal elastic constants.

The high-entropy carbides are predicted as accurately as the binary carbides, with only approximately 0.5% variation between SQS configurations. Chemical complexity therefore introduces no apparent accuracy penalty for the systems studied. Accuracy instead varies with the transition-metal group: group-IV carbides are generally described well, whereas group-V carbides, particularly TaC, show larger errors in shear stiffness. Overall, these results demonstrate that pretrained uMLIPs can predict finite-temperature elastic properties of refractory carbides without additional training, typically within 5 - 15% of AIMD and at substantially lower computational cost.

# 5. Declarations

## 5.1 Funding

This work was supported by the Grant Agency of the Czech Technical University in Prague [grant No. SGS24/121/OHK2/3T/12]

## 5.2 Acknowledgments

Computational resources were provided through the project "Development of infrastructure facilities for PhD programmes at CTU" (registration no. CZ.02.01.01/00/22_012/0006347), supported by the Johannes Amos Comenius Programme.

# 6. Data availability

The ICET interface for generation of SQSs can be accessed from GitHub: github.com/bracerino/icet-sqs-gui. The input structures, calculation scripts, MD results, SQS generation records and analysis scripts are available on Zenodo at doi.org/10.5281/zenodo.23437746.

# 7. Author contributions

M.L. and J.D. conceived the original idea. M.L. wrote the initial draft, developed all used codes, performed all calculations, and analyzed the results presented in the article. J.D. and P.V. supervised the project and contributed to the execution of the calculations, analysis of results, and improvement of the manuscript. Š.S. improved the manuscript readability and assisted in results evaluation.

## 8. Competing interest

The authors declare no competing interests.

## 9. Declaration of Generative AI and AI-Assisted Technologies in the Writing Process

During the preparation of this manuscript, the authors used GPT-6 Astra and Claude Opus 5.5 for improving the readability, clarity, and grammatical quality of the written text. After using this tool, the authors carefully reviewed and edited the content as necessary.

## Supplementary Information

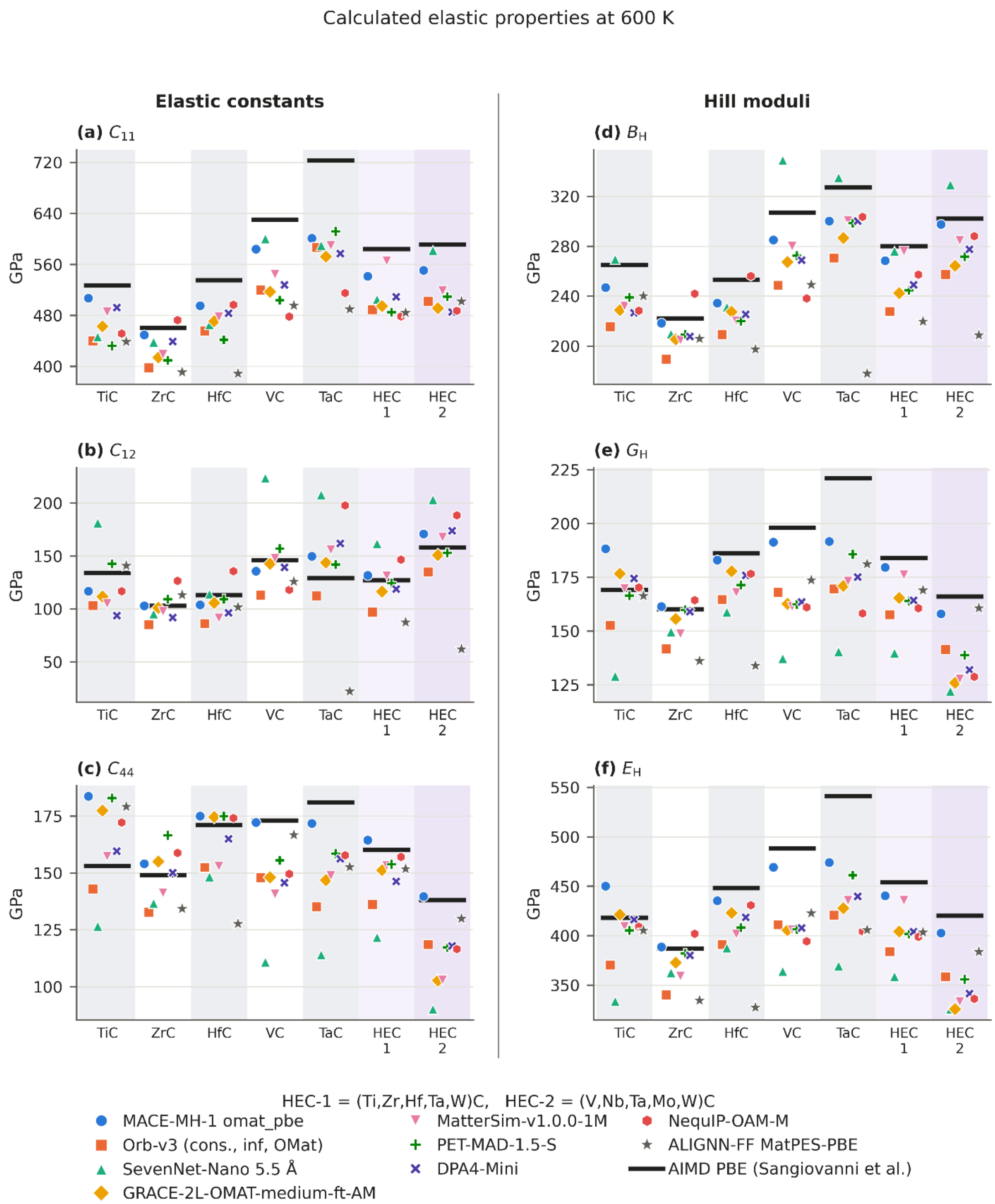

**Fig. S1:** Elastic constants $C_{11}$, $C_{12}$ and $C_{44}$ and Hill-averaged bulk ($B_H$), shear ($G_H$) and Young's ($E_H$) moduli of the 7 carbides at 600 K from the uMLIPs (symbols) compared with AIMD (black lines, Sangiovanni et al[10]).

Calculated elastic properties at 900 K

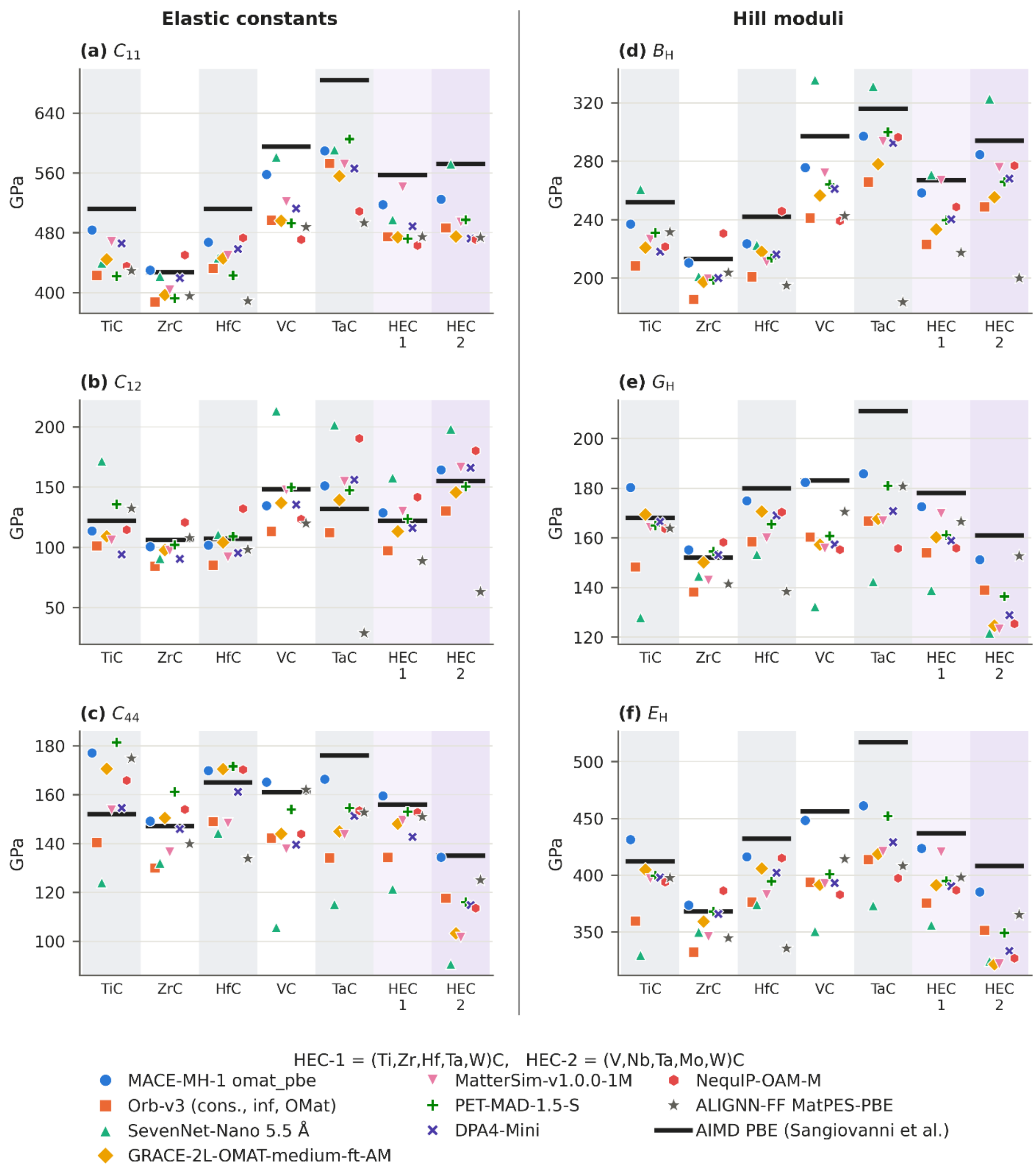

**Fig. S2:** Elastic constants $C_{11}$, $C_{12}$ and $C_{44}$ and Hill-averaged bulk ($B_H$), shear ($G_H$) and Young's ($E_H$) moduli of the 7 carbides at 900 K from the uMLIPs (symbols) compared with AIMD (black lines, Sangiovanni et al[10]).

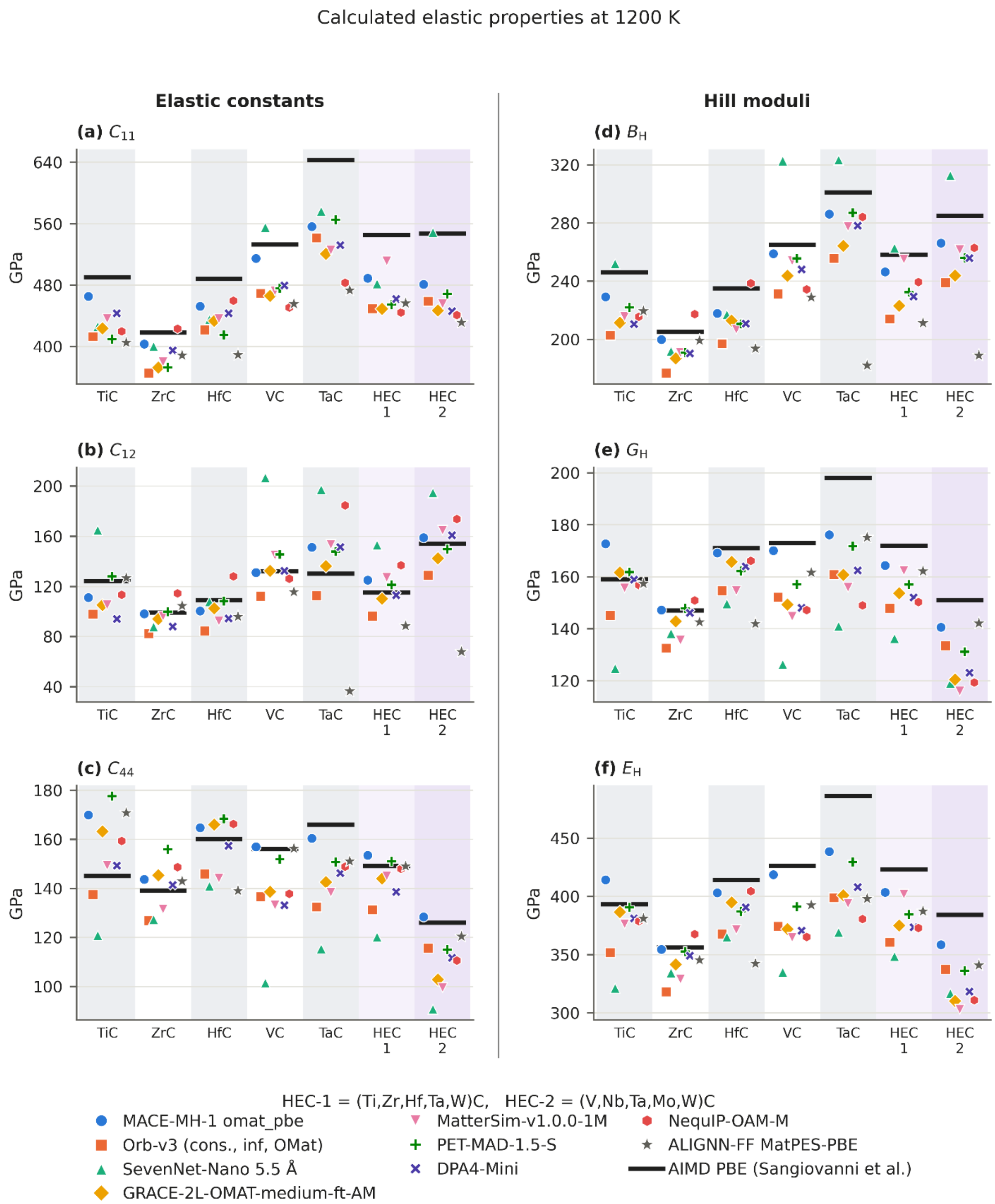

**Fig. S3:** Elastic constants $C_{11}$, $C_{12}$ and $C_{44}$ and Hill-averaged bulk ($B_H$), shear ($G_H$) and Young's ($E_H$) moduli of the 7 carbides at 1200 K from the uMLIPs (symbols) compared with AIMD (black lines, Sangiovanni et al[10]).

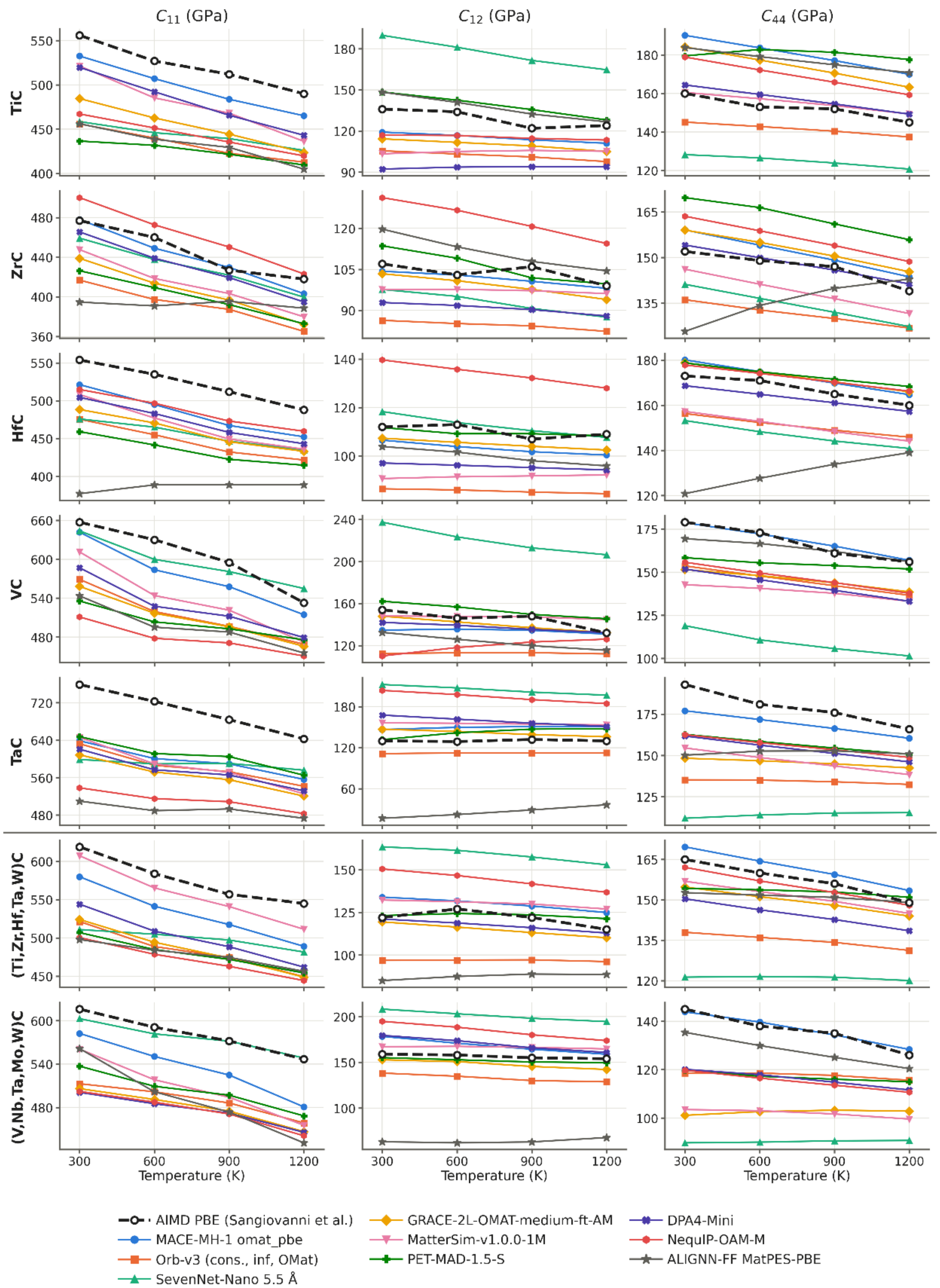

**Fig. S4:** Temperature dependence of the elastic constants $C_{11}$, $C_{12}$ and $C_{44}$ predicted by the nine uMLIPs for the seven investigated carbides. Dashed black lines show the AIMD reference values of Sangiovanni et al[10]. For the high-entropy carbides, predictions are averaged over six SQS configurations.

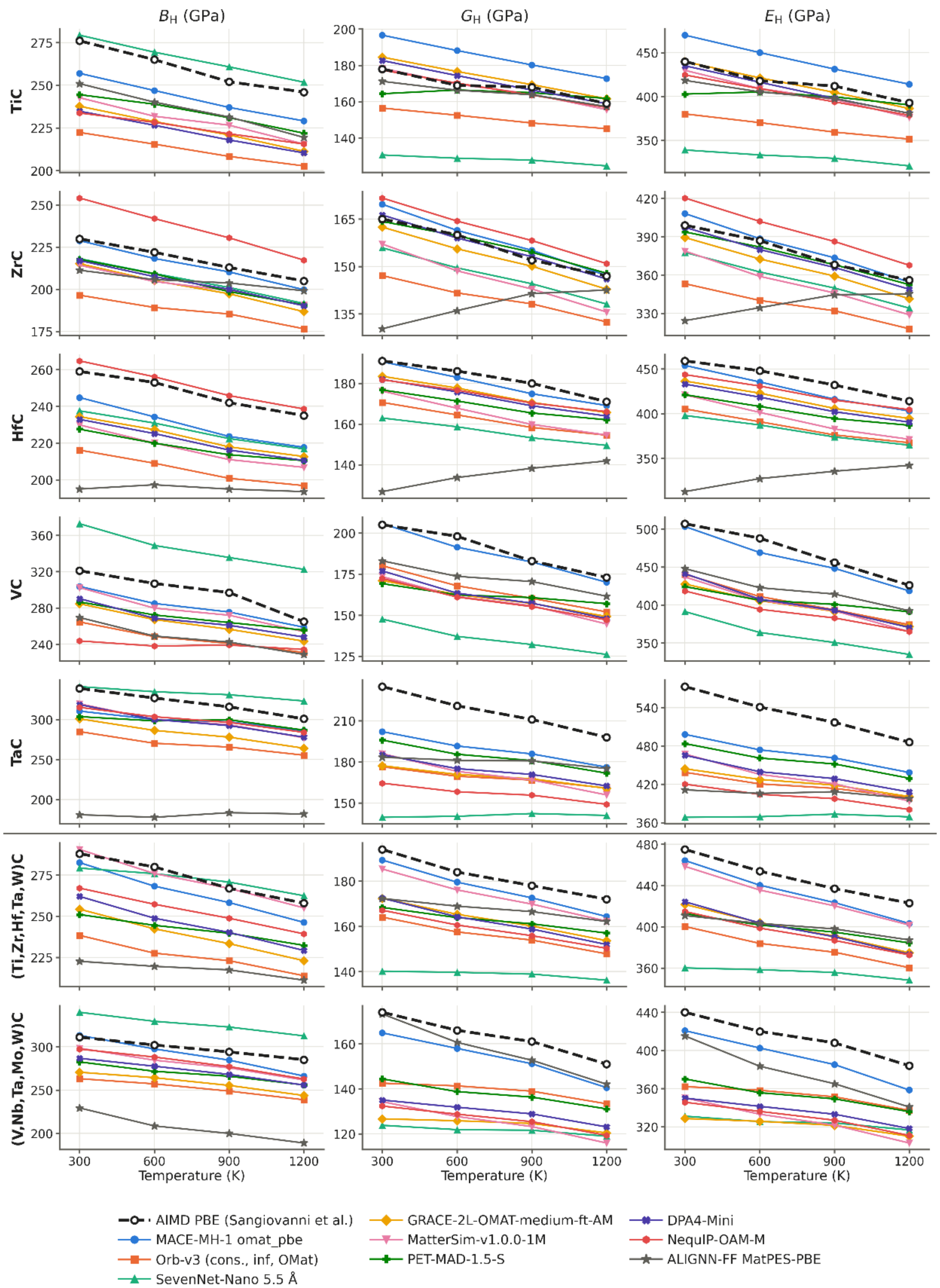

**Fig. S5:** Temperature dependence of the bulk ($B_H$), shear ($G_H$), and Young's ($E_H$) moduli derived using Voigt–Reuss–Hill averaging for the seven investigated carbides and nine uMLIPs. Dashed black lines show the AIMD reference values of Sangiovanni et al[10]. For the high-entropy carbides, predictions are averaged over six SQS configurations.